\documentstyle [preprint,aps]{revtex}
\begin{document}
\preprint{FSUHEP-950802}{MSUHEP-50801}{CTEQ-95/501}
\draft
\title{Event Structure in the Production of High Mass Photon Pairs}
\author{J. Huston}
\address{Physics Department, Michigan State University, East Lansing,
Michigan 48827}
\author{J.F. Owens}
\address{Physics Department, Florida State University, Tallahassee, Florida
32306-3016}
\date{\today}
\maketitle
\begin{abstract}
The event structure associated with the production of massive photon pairs is
studied to ${\cal O}(\alpha^2\alpha_s)$. The cross sections for two photon and
two photon plus one jet events are given. The Dalitz plot structure of events
with two photons and a jet is also
presented. These distributions will provide checks on our understanding of the
production mechanisms for massive photon pairs, a signal thought to be
important for searches for the Higgs boson in the intermediate mass region.
\end{abstract}
\pacs{ }
\section{Introduction}
\label{intro}
The production of photons in high momentum transfer collisions is useful for a
variety of reasons \cite{JFO}. The single photon invariant cross section is
one of
several important sources of information on the gluon distribution through the
role played by the subprocess $qg\rightarrow \gamma q$ \cite{OT}. The photon
plus jet
cross section may also provide additional constraints as the quantity of such
data is increased. Photon pair production has already played a role in
confirming the quark charge assignments and is one of many large momentum
transfer processes used to check the QCD-based description of high-p$_T$
scattering \cite{BBF}. Most recently, photon pair production has received
attention
because of its role in the search for the Higgs boson in the intermediate mass
region \cite{GKW}. The signature provided by the decay $H\rightarrow \gamma
\gamma$ is
particularly clean, enabling one, in principle, to detect the relatively rare
processes of Higgs production. However, a crucial element in this search is to
understand the two-photon background that results from conventional scattering
processes \cite{BO}. In addition, it has been suggested that the observation of
the
production of photon pairs may serve as a useful signal for new physics
processes\cite{Rizzo}.

The purpose of this paper is to present predictions for the event structure in
two photon events. In particular, we will focus on events with one additional
jet in the final state. Such events can be conveniently analyzed in terms of
several angular distributions and the structure of a two-dimensional Dalitz
plot. In addition, the presence of the jet allows one to define three classes
of events depending on whether the jet has the most, second most, or least
transverse energy when compared to the two photons. A similar analysis has been
proposed for events with one photon and two jets in the final state\cite{KO}.

The plan of the paper is as follows. In Sec.\ \ref{sec:program} a brief
description of the program will be given. In Sec.\ \ref{sec:plots} the
characteristics of the two photon plus jet final states will be presented for a
set of cuts which could be appropriate for either the CDF or D0 experiments.
Our summary and conclusions will be presented in Sec.\ \ref{sec:summary}.

\section{${\cal O}(\alpha^2 \alpha_s)$ Calculation}
\label{sec:program}

The program used for this analysis has been described in Ref.\ \cite{BOO},
so only a brief summary will be presented here. To order ${\cal O}(\alpha^2)$
the production of photon pairs proceeds via $q\bar q \rightarrow \gamma
\gamma$. In addition, there are contributions from final states with a
single photon and a jet where the jet gives rise to a photon. This is
estimated by including the $q\bar q \rightarrow \gamma g \text{ and }
qg\rightarrow \gamma q$ subprocesses convoluted with the appropriate
photon fragmentation function. The latter is formally of ${\cal O}(\alpha
/\alpha_s)$ which yields an overall factor of $\alpha^2$ when combined
with the factor of $\alpha \alpha_s$ coming from the subprocess. Similarly,
one can have final states with two jets, each of which fragments to a photon
plus hadrons. This contribution is given by the sum of all $2\rightarrow2$
parton-parton scattering subprocesses convoluted with two fragmentation
functions. These last two contributions are referred to as single and
double bremsstrahlung processes, respectively. Although formally of ${\cal O}
(\alpha^2 \alpha_s^2)$, one can also have significant contributions from the
subprocess $gg\rightarrow \gamma \gamma$ at high energies where small values
of $x$ become important since the gluon distribution is large enough there to
compensate for the suppression from the running coupling.

The ${\cal O}(\alpha^2 \alpha_s)$ next-to-leading-logarithm calculation
includes contributions from the one-loop corrections to $q\bar q \rightarrow
\gamma \gamma$ and the $2\rightarrow 3$ subprocesses $q\bar q \rightarrow
\gamma \gamma g \text{ and } qg\rightarrow \gamma \gamma q$. The actual
calculation is performed using a combination of Monte Carlo and analytic
integration techniques. The regions of phase space corresponding to soft or
collinear singularities are isolated by the use of cutoffs, $\delta_s
\text{ and } \delta_c$, applied to the Mandelstam variables for the
$2\rightarrow 3$ subprocesses. The squared matrix elements are integrated over
the singular regions using dimensional regularization. The soft singularities
cancel with corresponding terms from the one-loop corrections. Finally, the
collinear singularities are factorized using the ${\rm \overline{MS}}$ scheme
and
absorbed into the corresponding distribution or fragmentation function.
All remaining integrations are performed via Monte Carlo. When properly
executed, the cutoff dependence in such a calculation cancels between the
two-body and three-body contributions which is appropriate since the cutoffs
merely serve to mark where one switches from a Monte Carlo to an
analytic integration technique. Additional details can be found in Ref.\
\cite{BOO}.

The calculation outlined here includes contributions from all ${\cal
O}(\alpha^2 \alpha_s)$ subprocesses where the photons both are part of the
hard scatter. There are also fragmentation contributions where the hard
scattering process is calculated to ${\cal O}(\alpha_s^3)$ and two of the
jets fragment to photons. These contributions have not been incorporated into
the calculation. However, for collider energies isolation cuts are employed to
help reduce the hadronic background to the photon signals. These isolation cuts
greatly reduce the fragmentation contributions. Another point to remember is
that this calculation is
next-to-leading-order for observables involving the photon pairs. However, for
the two photon plus jet final states only the $2\rightarrow 3$ subprocesses
contribute, so this represents a leading order calculation for that case.
There are, for example, corrections from $2\rightarrow 4$ subprocesses where
one of the final jets is not detected. These contributions are beyond the scope
of this calculation.

\section{Predictions}
\label{sec:plots}

Predictions for two photon and two photon plus jet final states were
generated using the program described in the previous section. The CTEQ2M
distributions \cite{CTQLNG} were used with the factorization and
renormalization scales chosen to be the maximum $E_T$ in each event. A
minimum transverse momentum cut of 10 GeV is applied to the photons and the
jet. In addition, the pseudorapidities of the two photons are required to
satisfy $\vert \eta \vert
\leq 1$ and that of the jet to satisfy $\vert \eta \vert \leq 3$.
Both photons are required to be isolated, {\it i.e.}, to have less than
4 GeV of additional energy in a cone of radius $R = \sqrt{\Delta \eta^2
+\Delta \phi^2}$ of 0.7 centered on the photon direction. Effectively, this
just requires the two photons and the jet to be separated by more than
$\Delta R=0.7$. Such a cut greatly reduces contributions from subprocesses
involving photon fragmentation functions. In the experimental analysis the
isolation cut also discriminates against jets fragmenting into a leading
$\pi^0 \ {\rm or\ }\eta$, and thus greatly decreases the background to the
true two photon signal. These cuts are typical of what would be appropriate
for an analysis by either the D0 or CDF groups.

In Fig.\ \ref{etxsec} the cross section is shown for two photon and two photon
plus jet production as a function of the total transverse energy
($E_T^{\gamma \gamma}$) of the two photons. The photon and jet (if present)
are subject to the cuts
described above. The top curve shows the total cross section while the bottom
curve shows the cross section with a jet satisfying the above cuts.
The fraction of the cross section coming from events
with a jet in the final state that satisfies these cuts varies from about 60\%
for $E_T^{\gamma \gamma} = 100$ GeV to about 30\% at the upper end of the
curve. This ratio is larger than might have been expected since the rapidity
interval for the accompanying jet is much larger than that for the photons.

In Fig. \
\ref{subproc} the fractions of the two photon plus jet cross section coming
from the $q\bar q\ {\rm and\ } qg$ subprocesses are shown versus the total
transverse energy in the final state, denoted by $\Sigma E_T$. At low $\Sigma
E_T$ the $qg$ subprocess dominates while the $q\bar q$ subprocess dominates
at high $\Sigma E_T$ values. This is primarily due to the relative sizes of
the gluon and quark distributions at low and high values of $x$.

For those events with a jet in the final state, the fractions of the cross
section are plotted
in Fig. \ \ref{etjet} according to whether the jet has the largest, second
largest, or least $E_T$ among the two photons and the jet. It is useful to
classify the events in this manner since the jet is the one object
distinguishable form the others. The photon was used in this manner in a
similar analysis of photon plus two jet events in Ref.\ \cite{KO}. The results
in this figure show that at high $\Sigma E_T$ it is most likely that the jet
will have less $E_T$ than either of the photons, while at lower values it is
more likely that the jet will have more $E_T$ than one or both of the photons.

For events with two photons and a jet in the final state, four kinematic
variables in addition to the three-body center-of-mass energy must be specified
in order to determine the final state configuration. It is convenient to use
scaled energy variables for two of these. In the parton-parton center-of-mass
frame the energies of the final state particles can be labelled as $E_3,
E_4, \ {\rm and \ } E_5$ in decreasing order of energy. The energy fractions
$x_i
=2 E_i/E_{tot}$ can then be constructed, with $E_{tot}=E_3+E_4+E_5$. Hence,
$x_3 + x_4 + x_5 =2$. The variable $x_3$ satisfies $\frac 2 3 \leq x_3 \leq 1$,
where the lower limit corresponds to the symmetric ``Mercedes Benz''
configuration with all three $x_i's$ being equal. As $x_3$ tends towards one,
the vectors of particle 4 and 5 become collinear and
opposite to that of particle 3. The allowed range for $x_4$ is $0.5 \leq
x_4 \leq 1$ where the lower limit corresponds to $x_4=x_5$ with particle 4 and
5 being collinear while the upper limit corresponds to $x_5=0 \text{ and }
x_3+x_4=1$. Only two of the $x_i's$ are independent; these will be taken to be
$x_3 \text{ and } x_4$. Two angles are needed to complete the description of
the final state. In the parton-parton
center-of-mass frame, let $\theta^*$ be the angle between the direction of the
incoming proton and the
direction of the jet or photon with energy $E_3$. The plane containing
the beam and this jet or photon will be referred to as the scattering plane.
The second angle is $\psi$ which is the angle between the normal to the
scattering plane and the normal to the plane containing the remaining jet or
photons.

The main point of this analysis is that examination of various distributions
in terms of these kinematic variables may provide useful information concerning
the accuracy of the current QCD-based description of two photon production.
For example, some lego plots of $x_3 \text{ versus } x_4$ for events where the
jet is the most energetic are shown in Fig.\ \ref{jet1-60}, \ref{jet1-200}, and
\ref{jet1-400}. In these cases the
two photons recoil against the jet. The $p_T, \eta,$\ and isolation cuts
described above have been applied. In Fig.\ \ref{jet1-60} a cut requiring the
total $E_T$ to be greater than 60 GeV has been applied. No obvious structure
is visible. However, any structure is hidden because the total $E_T$ cut is
low with respect to the individual minimum $E_T$ requirements. In Fig.\
\ref{jet1-200} the cut on the total $E_T$ has been raised to 200 GeV and in
Fig.\
\ref{jet1-400} to 400 GeV. As the minimum for the total $E_T$ is raised an
enhancement in the region $x_3, x_4 \rightarrow 1$ begins to appear, reflecting
the existence of a pole in the matrix element at $x_3=x_4=1$. In this region
one of the photons becomes soft, {\it i.e.,} approaches the minimum $E_T$
value of 10 GeV. For high values of the total $E_T$ it is expected that the
majority of events in this class would have a high $E_T$ jet, a slightly less
energetic photon, and a relatively soft second photon.

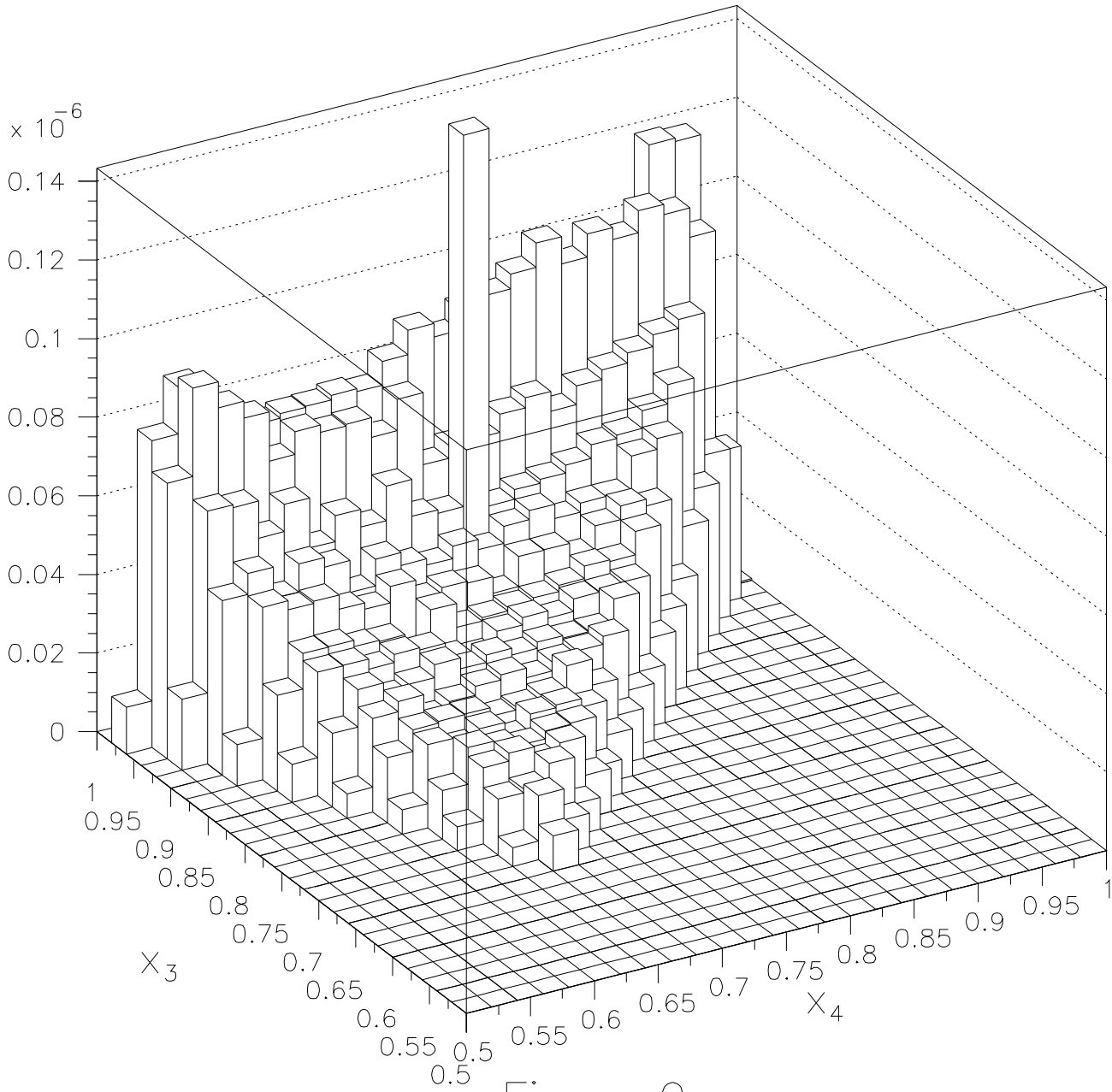
\begin{figure}
\caption{The same as the previous figure except for the $q g$
initial state. \hbox{\ \ \ \ \ \ \ \ \ \ \ \ \ \ \ \ \ \ \ \ \ \ \ \ }
\label{qg}}
\end{figure}

The lego plots for the case where the jet is the second most energetic show a
similar structure. A related, but somewhat different, pattern emerges when the
jet has less energy than either of the photons. For this configuration
the jet and the photon with the least energy of the pair recoil against the
most energetic photon. Again, a strong pole structure becomes increasingly
evident for progressively higher values of total $E_T$. However, as shown in
Fig.\ \ref{jet3-400}, the shape
of the emerging pole structure is somewhat broader in this case as compared to
the case where the jet is the most energetic. This can be ascribed to the fact
that the subprocess $qg \rightarrow \gamma \gamma q$ does not have a
soft singularity when the quark energy goes to zero, but it does when one of
the photon energies goes to zero. This can be further illustrated by comparing
Fig.\ \ref{qqbar} and Fig.\ \ref{qg} which show the contributions to Fig.\
\ref{jet3-400} from the $q\bar q \rightarrow \gamma \gamma g \text{ and }
qg\rightarrow \gamma \gamma q$ subprocesses, respectively.

We have examined the $\cos(\theta^*) \text{ and } \psi$
distributions for the three cases where the jet has the most, second most, or
least energy with minimum total $E_T$ cut of 200 GeV applied. The $\psi$
distributions are essentially the same for all three cases while some minor
differences are apparent in the $\cos(\theta^*)$ distributions. However, the
exact shapes are rather cut dependent and so have not been shown. The
appropriate distributions can easily be generated once the appropriate cuts
have been specified.

Several two photon events have been observed by the CDF Collaboration with
values of total $E_T$ exceeding 400 GeV \cite{CDF}. Two of the three events
have a jet associated with the two photons. In one event the jet
is the most energetic and for the other it is the least energetic. Based on
the previous discussion one might expect that the photon or jet with the least
energy might have an $E_T$ value near the minimum allowed, {\it i.e.,} that the
event would be in the region of the corresponding lego plot where the
enhancement due to the pole occurs. Instead, the energies are closer than would
have been anticipated. The unusual topologies of these events may simply be
due to statistical fluctuations, but the continued presence of a large number
of such events as the data sample grows could signal the need for additional
sources of high mass photon pairs in the calculation.

\section{Summary and Conclusions}
\label{sec:summary}

As preparations for the next generation of colliders continues, interest in
the $\gamma \gamma$ channel has remained high since it can be used to search
for the intermediate mass Higgs boson. It is, therefore, necessary to
understand the conventional QCD production mechanisms for photon pairs in
order to be able to provide reliable background estimates. The study of the
event structure of two photon plus jet events using the distributions suggested
in this paper may well provide useful tests of our understanding.

\begin{figure}
\caption{The two photon cross section (upper curve) and the two photon plus
jet cross section (lower curve) versus the total transverse energy of the
two photons, $E_T^{\gamma \gamma}$, using the cuts
discussed in the text. \label{etxsec}}
\end{figure}

\begin{figure}
\caption{The fractions of the two photon plus jet cross section coming from
the $q\bar q \text{(solid) and the } qg$ subprocesses \text{(dashed)}.
\label{subproc}}
\end{figure}

\begin{figure}
\caption{The fractions of the two photon plus jet cross section where the
jet has the highest (solid), second highest (dashed), or least (dotted) $E_T$.
\label{etjet}}
\end{figure}

\begin{figure}
\caption{Lego plot of $x_3$ versus $x_4$ when the jet has the largest of
the three $E_T$ values and $\Sigma E_T$ is required to be greater than 60 GeV.
\label{jet1-60}}
\end{figure}

\begin{figure}
\caption{Lego plot of $x_3$ versus $x_4$ when the jet has the largest of
the three $E_T$ values and $\Sigma E_T$ is required to be greater than 200
GeV. \label{jet1-200}}
\end{figure}

\begin{figure}
\caption{Lego plot of $x_3$ versus $x_4$ when the jet has the largest of
the three $E_T$ values and $\Sigma E_T$ is required to be greater than 400
GeV.
\label{jet1-400}}
\end{figure}

\begin{figure}
\caption{Lego plot of $x_3$ versus $x_4$ when the jet has the least of
the three $E_T$ values and $\Sigma E_T$ is required to be greater than 400
GeV. \label{jet3-400}}
\end{figure}

\begin{figure}
\caption{The contribution to the plot in the previous figure from the $q\bar q$
initial state. \hbox{\ \ \ \ \ \ \ } \label{qqbar}}
\end{figure}

\begin{references}

\bibitem{JFO} J.F. Owens, Rev. Mod. Phys. {\bf 59}, 465 (1987).

\bibitem{OT} J.F. Owens and W.-K. Tung, Annu. Rev. Nucl. Part. Sci. {\bf 42},
291 (1992).

\bibitem{BBF} E.L. Berger, E. Btaaten, and R.D. Field, Nucl. Phys. {\bf B239},
52 (1984).

\bibitem{GKW} J.F. Gunion, G.L. Kane, and J. Wudka, Nucl. Phys. {\bf B299}, 231
(1988).

\bibitem{BO} H. Baer and J.F. Owens, Phys. Lett. {\bf B205}, 377 (1988).

\bibitem{Rizzo} T. Rizzo, Phys. Rev. {\bf D51}, 1064 (1995).

\bibitem{KO} S. Keller and J.F. Owens, Phys. Lett. {\bf B269}, 445 (1991).

\bibitem{BOO} B. Bailey, J. Ohnemus, and J.F. Owens, Phys. Rev. {\bf D46}
2018, 1992.

\bibitem{CTQLNG} H.-l Lai {\it et al.}, Phys. Rev. {\bf D51}, 4763 (1995).

\bibitem{CDF} R. Blair {\it et al.} (the CDF Collaboration), Proceedings of
the 10th Topical Workshop on Proton-Antiproton Collider Physics, Fermilab,
May, 1995.

\end{references}
\end{document}